\documentclass[12pt,preprint]{aastex}

\slugcomment{ApJ in press (2003, v594)}

\shorttitle{GRBs from Neutron Star Kicks}
\shortauthors{Huang et al.}

\begin{document}



\title{Gamma-Ray Bursts from Neutron Star Kicks}


\author{Y. F. Huang\altaffilmark{1,2}, Z. G. Dai\altaffilmark{1}, 
   T. Lu\altaffilmark{1}, K. S. Cheng\altaffilmark{3}, 
   and X. F. Wu\altaffilmark{1}}






\altaffiltext{1}{Department of Astronomy, Nanjing University, Nanjing 210093, China
     (hyf@nju.edu.cn)}
\altaffiltext{2}{LCRHEA, IHEP, Chinese Academy of Sciences, Beijing 100039, China}
\altaffiltext{3}{Department of Physics, The University of Hong Kong, Pokfulam Road,
    Hong Kong, China}


\begin{abstract}
The idea that gamma-ray bursts might be a kind of phenomena 
associated with neutron star kicks was first proposed by Dar 
\& Plaga (1999). Here we study this mechanism in more detail
and point out that the neutron star should be a high speed
one (with proper motion larger than $\sim 1000$ km/s). It is 
shown that the model agrees well with observations in many aspects,
such as the energetics, the event rate, the collimation, 
the bimodal distribution of durations, the narrowly clustered
intrinsic energy, and the association of gamma-ray bursts with 
supernovae and star forming regions. We also discuss the 
implications of this model on the neutron star kick mechanism, 
and suggest that the high kick speed were probably acquired 
due to the electromagnetic rocket effect of a millisecond 
magnetar with an off-centered magnetic dipole. 
\end{abstract}




\section{Introduction}

Gamma-ray bursts (GRBs), first detected serendipitously in 1967 
(Klebesadel, Strong \& Olson 1973), 
are intense $\gamma$-ray flashes lasting for tens of seconds that occur 
randomly in the deep sky. The great debate on the distances of GRBs once
lasted for about 30 years. The problem was finally resolved in 1997,
when X-ray, optical, and radio afterglows from some GRBs began to be
discovered due to the successful operation of the Italian-Dutch
BeppoSAX satellite (van Paradijs et al. 1997; Costa et al. 1997). 
Observations on GRB afterglows in the past six years have definitely 
shown that at least most long GRBs are of cosmological origin. Under 
isotropic assumption, GRBs will be the most powerful 
explosions in the Universe since the Big Bang (Kulkarni et al. 1999; 
Andersen et al. 1999). The famous 
fireball model, which incorporates internal shocks to account for the 
main bursts, and external shocks to account for afterglows, becomes 
the most popular model (Piran 1999; van Paradijs, Kouveliotou \& 
Wijers 2000; M\'esz\'aros 2002).  However, the nature of 
GRB ``central engines'' is still far from clear and is still one of 
the greatest mysteries in modern astrophysics.

Currently, one popular class of ``engines'' 
involves the core collapses of very massive stars (heavier than $\sim 40
M_{\odot}$), often referred to as hypernovae or collapsars 
(Paczy\'nski 1998; Fryer, Woosley \& Hartmann 1999).  
However, the core collapse is a very complicated process. Without further 
careful simulations, it is still largely unclear whether hypernovae and 
collapsars can successfully generate the required ultra-relativistic 
ejecta as expected. Another major class of candidates involves the merger of 
two compact stars, such as neutron star binaries or neutron star-black hole 
binaries(e.g., Goodman, Dar \& Nussinov 1987; Eichler et al. 1989; Narayan, 
Paczy\'nski \& Piran, 1992; Bloom, Sigurdsson \& Pols 1999). But these 
mergers are usually outside of star forming regions, and additionally 
they have difficulties in accounting for the long durations of most GRBs.

In 1999, Dar \& Plaga (1999; also see Dar 1999) discussed the possibility 
that GRBs might come from neutron star kicks. They suggested that the 
natal kick of a neutron star is due to the emission of a relativistic jet
from the compact object.  Momentum conservation then 
indicates that the kinetic energy enclosed in the jet is 
$\sim 4 \times 10^{51}$ ergs (Dar \& Plaga 1999; Dar 1999),
enough to account for typical GRBs.  Largely 
based on this assumption, they have proposed the cannonball 
model of GRBs (see Dado, Dar \& De R\'ujula 2002a and references
therein). 

In this research, we study the energy mechanism suggested by Dar et al.
in more detail. We show that the model can naturally meet many of the 
requirements imposed by GRB observations. We specially point out that the 
neutron star in this model should be a high speed one ($> 1000$ km/s), 
which probably receives the large kick velocity through the electromagnetic 
rocket effect.

\section{Momentum Conservation}

Observations of GRBs and their afterglows have provided useful clues on 
the nature of GRB central engines. In the currently popular models of 
GRBs the central engine must satisfy the following requirements: 
(i) The central engine should release an isotropic-equivalent energy 
of $\sim 10^{51}$ --- $10^{53}$ ergs. (ii) The energy release should 
usually be highly collimated, with typical half opening angle of 
$\theta \sim 0.1$ radian (e.g., Frail et al. 2001). (iii) There should 
be very few baryons in the beamed ejecta, so that it can move 
ultra-relativistically with a bulk Lorentz factor 
of $\gamma \geq 100$ --- 1000 (Lithwick \& Sari 2001). 
(iv) The progenitors should be embedded in star forming galaxies and 
should follow the cosmic star formation rate (Wijers et al. 1998; 
Fruchter et al. 1999). In fact, there is accumulating evidence that long 
duration GRBs are associated with supernovae of Type Ic (Kulkarni et al. 
1998; Galama et al. 1998; Bloom et al. 1999; Reeves et al. 2002). Recent 
good proofs for this GRB-supernova connection come from the observations 
of afterglows from GRBs 020405 and 030329 (Price et al. 2003; Masetti et 
al. 2003; Stanek et al. 2003). (v) The event rate 
should be $\sim 10^{-5}$ --- $10^{-4}$ per typical galaxy per year, 
taking into account the beaming effects. The rate is estimated as 
$\sim 10^{-7}$ --- $10^{-6}$ per typical galaxy per year under 
isotropic assumption. (vi) The life time of the central engines should 
be $\sim 10$ --- 100 s, during which the energy release should be highly 
variable (Sari \& Piran 1997; Kobayashi, Piran \& Sari 1997). 

On the other hand, radio pulsars are observed to have a mean 
three-dimensional velocity of 200 --- 500 km/s, with a significant 
population having velocities greater than 1000 km/s (Frail, Goss \& 
Whiteoak 1994; Cordes \& Chernoff 1998). Since the average space velocity 
of normal stars in the Milky Way is only about 30 km/s, it is generally 
believed that pulsars must receive a substantial ``kick'' at birth 
(van Den Heuvel \& van Paradijs 1997; Spruit \& Phinney 1998; Lai, 
Chernoff \& Cordes 2001). Neutron star kick is surely a catastrophic 
and violent process in the Universe. 
While the result of the process (i.e., large 
proper motions of neutron stars) has been definitely observed, it seems 
that we still do not detect any phenomena that are directly connected to 
the kick process itself. Note that supernovae are still not the specific 
phenomena that we are talking about, since we can never predict whether 
a high speed neutron star has been produced or not just from the supernova 
observations. Here we suggest that the mysterious GRBs are just the 
interesting phenomena that we are seeking for, i.e., the emergence of a
GRB may indicate the birth of a high speed neutron star.

The possible intrinsic connection between GRBs and neutron star kicks 
was first realized by Dar \& Plaga (1999; also see Dar 1999). 
They assumed that a relativistic jet is 
responsible for the large kick velocity of pulsars. 
The jet then potentially has the ability
to account for a GRB. Denoting the mass of the high speed neutron star 
as $M_{\rm NS}$ and its kick velocity as $V_{\rm NS}$, the total energy 
($E_{\rm flow}$) enclosed in the recoiling outflow can be easily 
calculated from momentum conservation (Dar \& Plaga 1999; Dar 1999),  
\begin{equation}
E_{\rm flow} = M_{\rm NS} V_{\rm NS} c = 8.3 \times 10^{51} \;
  {\rm ergs} \; \cdot \left( \frac{M_{\rm NS}}{1.4 M_\odot} \right)
  \cdot \left( \frac{V_{\rm NS}}{1000 \, {\rm km/s}}\right),
\label{Eflow1}
\end{equation}
where $c$ is the speed of light.  However, usually not all of this energy can 
be used to power a GRB. Assuming that a portion $\epsilon$ of $E_{\rm flow}$ 
is deposited into electron-positron pairs, and that they are beamed into a 
cone with a small half opening angle of $\theta$, then an on-axis observer 
will detect an intense GRB with an isotropic equivalent energy of
\begin{equation}
E_{\rm iso} = \frac{2 \epsilon E_{\rm flow}}{1 - \cos \theta}
   \approx 4 \epsilon M_{\rm NS} V_{\rm NS} c \theta^{-2} \\
   = 3.3 \times 10^{53} {\rm ergs} \cdot \left(\frac{\epsilon}{0.1}
   \right) \cdot \left(\frac{\theta}{0.1}\right)^{-2}
   \cdot \left( \frac{M_{\rm NS}}{1.4 M_\odot} \right)
   \cdot \left( \frac{V_{\rm NS}}{1000 \, {\rm km/s}}\right).
\label{Eiso2}
\end{equation}
For $\epsilon$ values as high as 0.3 and $\theta$ values as low as 0.05, 
$E_{\rm iso}$ can reach $\sim 4.0 \times 10^{54}$ ergs, enough to account for 
all the GRBs localized so far. We thus see that GRBs basically can be due to 
the birth of high speed neutron stars ($V_{\rm NS} \geq 1000$ km/s).

\section{Kick Mechanism}

To evaluate $\epsilon$ and $\theta$ more rationally, and to examine whether 
this mechanism can meet other observational requirements listed at the 
beginning of this section, we must resort to the 
detailed kick mechanism, which itself, however, is still a bit uncertain. 
According to the characteristics of the recoiling outflows, current kick 
models can be divided into three main categories, i.e., hydrodynamically 
driven kicks, neutrino driven kicks and electromagnetic radiation driven 
kicks (Lai, Chernoff \& Cordes 2001). We discuss them one by one below. 

In hydrodynamically driven kick mechanisms, asymmetric matter ejection 
and/or asymmetric neutrino emission due to global asymmetric perturbations 
of pre-supernova cores is involved (Janka \& M\"uller 1994; Burrows \&
Hayes 1996). The timescale of the kick process has been estimated as 
$\tau_{\rm kick} \sim 0.1$ s (Lai, Chernoff \& Cordes 2001). 
However, it turns out to be very unlikely that these mechanisms are able 
to account for the observed pulsar velocities in excess of about 500 
km/s (Janka \& M\"uller 1994). They should be irrelevant to the high 
speed neutron stars that interest us here. 

In neutrino driven kick mechanisms, asymmetric neutrino emission induced 
by strong magnetic fields acts as the working medium of the rocket effect. 
There are mainly two kinds of detailed mechanisms. In the first mechanism, 
since the cross section for $\nu_{\rm e}$  ($\bar \nu_{\rm e}$) 
absorption on neutrons (protons) depends on 
the local magnetic field strength, asymmetric neutrino emission can be 
produced if the field strengths at the two opposite poles of the neutron 
star are different. To generate a recoil velocity of $V_{\rm NS} \sim 300$
km/s would require that the difference in the field strengths at the two 
opposite stellar poles be at least $10^{16}$ G (Lai, Chernoff \& Cordes 
2001; Lai \& Qian 1998). 
The second mechanism relies on the effect of parity violation, which 
indicates that the neutrino opacities and emissivities in a strongly 
magnetized nuclear medium depend asymmetrically on the directions of neutrino 
momenta with respect to the magnetic field (Lai, Chernoff \& Cordes 2001; 
Arras \& Lai 1999). The 
resulting kick velocity is $V_{\rm NS} \sim 50 (B/10^{15} \, {\rm G})$ km/s. 
To generate a recoil velocity of 1000 km/s, the magnetic field should 
be $B \sim 2 \times 10^{16}$ G. Although evidence for the existence of 
magnetars with superstrong magnetic field approaching $10^{15}$ G has 
been revealed in soft gamma repeaters (SGRs) and anomalous X-ray pulsars 
(AXPs) (Thompson \& Duncan 1995; Kouveliotou et al. 1998; Hurley et al. 
1999; Ibrahim, Swank \& Parke 2003), 
a field strength of $B \geq 2 \times 10^{16}$ G at the 
neutron star surface is still unimaginably too large. The birth of high 
speed neutron stars should not be due to these mechanisms.  

Now we come to discuss the third class of kick mechanisms, electromagnetic 
radiation driven kicks. It has been shown that electromagnetic radiation 
from a rotating off-centered magnetic dipole imparts a kick to the neutron 
star (Harrison \& Tademaru 1975; Lai, Chernoff \& Cordes 2001). 
The kick comes at the expense of the spin kinetic 
energy. Under ``optima'' conditions, the maximum kick velocity can be 
(Lai, Chernoff \& Cordes 2001)
\begin{equation}
V_{\rm NS} \sim 1400 (R / 10 \, {\rm km})^2 (P / 1 \, {\rm ms})^{-2}
\; \; {\rm km/s},
\label{Vns3}
\end{equation}
where $R$ and $P$ are the radius and period of the neutron star 
respectively. Note that the rotational kinetic energy of a neutron star 
with a moment of inertia of $I$ is (Usov 1992),
\begin{equation}
E_{\rm spin} = \frac{1}{2} I \left( \frac{2 \pi}{P} \right)^2
   \approx 2 \times 10^{52} \left( \frac{I}{10^{45} \; {\rm g \cdot cm}^2} 
   \right) \cdot \left( \frac{P}{1 \, {\rm ms}} \right)^{-2} \;\; {\rm ergs},
\label{Espin4}
\end{equation}
enough to meet the requirement of Eq.~(\ref{Eflow1}). This electromagnetic 
rocket effect is usually considered as a ``postnatal'' kick, since for 
typical neutron stars with $B \sim 10^{12}$ G, the kick is attained on 
the initial spin-down timescale of $\tau_{\rm kick} \geq 10^9$ s. 
However, if the pulsar is a magnetar with a superstrong magnetic field, 
then the life time of the kick can be tens of seconds, i.e. 
(Usov 1992; Lai, Chernoff \& Cordes 2001),   
\begin{equation}
\tau_{\rm kick} \approx 50 \left(\frac{B}{3 \times 10^{15} {\rm G}}
\right)^{-2} \cdot \left(\frac{P}{1 \, {\rm ms}}\right)^2 \;\; {\rm s}.
\label{tau5}
\end{equation}
Since the existence of magnetars with superstrong magnetic field 
approaching $10^{15}$ G has been creditably proved from the studies of 
SGRs and AXPs (Thompson \& Duncan 1995; Kouveliotou et al. 1998; Hurley et al. 
1999; Ibrahim, Swank \& Parke 2003), 
we believe that electromagnetic radiation driven 
kick is the most viable mechanism responsible for the birth of high speed 
neutron stars. We will continue our analysis on the connection between 
GRBs and neutron star kicks in this frame work. 

\section{GRBs from Neutron Star Kicks}

Particle generation and acceleration at the surface of a millisecond 
magnetar have been studied in great detail by Usov (1992). Although the 
magnetic dipole involved here is off-centered, the process should largely 
be similar. As demonstrated by Usov, the component of electric field along 
magnetic field in the magnetosphere of a millisecond magnetar is extremely 
high. Plenty of electron-positron pairs are created directly due to the 
vacuum discharge (E $\rightarrow$ e$^+$ + e$^-$ + E) (Usov 1992). 
Additionally, pair creation through photon splitting ($\gamma$ + B 
$\rightarrow$ e$^+$ + e$^-$ + B) and photon-photon collision 
($\gamma$ + $\gamma$ $\rightarrow$ e$^+$ + e$^-$) may also play an 
important role in the process. Usov estimated that the fraction of the 
total spin-down energy that finally goes into electron-positron pairs, 
i.e., $\epsilon$ in our Eq.~(\ref{Eiso2}), is a few times 0.1.

Usov assumed that these energetic particles are emitted isotropically. 
This may deviate from the reality. According to pulsar theories 
(Ruderman \& Sutherland 1975; Cheng, Ho \& Ruderman 1986), 
particle generation and acceleration occur most likely 
at the polar cap or in a small region slightly above it. The emission of 
high energy particles thus should mainly be along the magnetic axis. In 
fact, the duty cycle (i.e., pulse width divided by period and then times 
360$^{\rm o}$) of radio pulsars is typically found to be 
$W_{\rm pulse} \sim 10^{\rm o}$, with a few exceptions where $W_{\rm pulse}$ 
can be as small as $\sim 3^{\rm o}$ or as large as tens of degrees 
(Manchester \& Taylor 1977). 
It is reasonable that the half opening angle of the primary electron-positron 
outflow should be less than $W_{\rm pulse}/2$. So, the $\theta$ parameter in 
our Eq.~(\ref{Eiso2}) can typically be evaluated as $\theta \sim 0.1$ radian, 
with the possibility that it can be as small as $\theta \sim 0.03$ radian 
in some cases. 

From the above analysis, we were convinced that GRBs can really be due to 
the kicks of high speed neutron stars. This model can naturally meet the
direct observational requirements listed in Sect. 2, for example: 
(i) The deposited energy is enough for GRBs. The isotropic equivalent 
energy can easily exceed $5 \times 10^{54}$ ergs. (ii) The collimation is 
safely guaranteed, with a typical beaming angle $\theta \sim 0.1$ . 
(iii) The ultra-relativistic motion (with Lorentz factor $\gamma \geq 100$ 
--- 1000) is reasonably expected, since the original outflows here are 
mainly composed of electrons and positrons. (iv) The model naturally 
explains the observed connection between GRBs and supernovae 
(for details, see Dado, Dar \& De R\'ujula 2002b, 2003), and the 
association of GRBs with star forming regions. (v) In this model, the 
durations of GRBs are obviously determined by the timescale of the kick 
process, which has been given in Eq.(~\ref{tau5}). It is in good agreement 
with observations. (vi) The model can also meet the requirement of GRB 
event rate. Let us have a look at this problem in some detail. The 
supernova rate in our Galaxy is $\sim$ 1/50 -- 1/30 per year (Tammann, 
L\"{o}ffler \& Schr\"{o}der 1994; van Den Bergh \& McClure 1994). Then 
the birth rate of neutron stars in a typical galaxy can be estimated 
as $\sim 10^{-2}$ per year. The percentage of high speed neutron stars 
is still a bit uncertain, but should be some value between 1\% and 10\% 
(Frail, Goss \& Whiteoak 1994; Cordes \& Chernoff 1998). 
So, the birth rate of high speed neutron stars is 
$\sim 10^{-4}$ --- $10^{-3}$ per galaxy per year. However, GRB emission 
from these objects is typically beamed into a small cone with a half 
opening angle $\theta \sim  0.1$ . After making compensation for the 
beaming effect, the predicted detectable GRB event rate should 
be $\sim 10^{-7}$ --- $10^{-6}$ per galaxy per year, just consistent 
with observations.  

The model also has the potential advantage of satisfying many other 
requirements inferred indirectly from GRB observations. For example,
the fast variability in GRB light curves indicates that internal shocks 
are preferable during the main GRB phase (Kobayashi, Piran \& Sari 1997). 
In our model, the possibility of generating internal shocks is greatly
increased thanks to the recently discovered apparent alignment 
of the spin axes and proper motion directions of the Crab and 
Vela pulsars (Caraveo \& Mignani 1999; Pavlov et al. 2000). This 
alignment indicates that the timescale of the kick should generally 
be much larger than the spin period of the neutron star, and that the 
velocity of the kicked material could make a non-zero angle $\Theta$ to 
the spin axis (Lai, Chernoff \& Cordes 2001). In other words, the GRB 
might come from a precessing jet (Fargion \& Salis 1995; Hartmann \& 
Woosley 1995; Blackman, Yi \& Field 1996; MacFadyen \& Woosley 1999;
Fargion 1999). 
In this case, Eqs.~(\ref{Eflow1}) and (\ref{Eiso2}) will become,
\begin{equation}
E_{\rm flow} = M_{\rm NS} V_{\rm NS} c / \cos \Theta,
\label{Eflow6}
\end{equation}
\begin{equation}
E_{\rm iso} \geq \frac{2 \epsilon M_{\rm NS} V_{\rm NS} c}
   {\theta \sin 2 \Theta}
   = \frac{1.7 \times 10^{52} \, {\rm ergs} }{\sin 2 \Theta}
   \cdot \frac{\epsilon}{0.1} \cdot \left(\frac{\theta}{0.1}
   \right)^{-1} \cdot \left(\frac{M_{\rm NS}}{1.4 M_\odot}\right) 
   \cdot \left(\frac{V_{\rm NS}}{1000 \; {\rm km/s}}\right),
\label{Eiso7}
\end{equation}
for $\theta \ll \Theta$ and $\theta \ll 1$. Eq.~(\ref{Eiso7}) means the 
GRB appears less powerful now, but the possibility that it can be 
detected increases by a factor of $\sim 4 \sin \Theta / \theta$. The 
precession of the jet may help to explain the rapid variability 
observed in GRB light curves (Roland, Frossati \& Teyssier 1994; 
Portegies-Zwart, Lee \& Lee 1999). 
We also notice that the space velocities of the Vela and Crab pulsars 
are not too large, i.e., $\sim 70$ --- 141 km/s and $\sim 171$ km/s 
respectively (Lai, Chernoff \& Cordes 2001). 
For high speed neutron stars, we can imagine that the $\Theta$ values 
should be very small so that Eq.~(\ref{Eiso2}) is still approximately 
applicable.

Frail et al. suggested that the $\gamma$-ray energy release in GRBs, 
corrected for geometry, is narrowly clustered around $5 \times 10^{50}$ ergs 
(Frail et al. 2001). It is interesting that our model strongly supports their 
conclusion. From Eq.~(\ref{Eflow1}) we see that the total energy enclosed 
in the recoiling outflow is $E_{\rm flow} \sim 8 \times 10^{51}$ ergs, 
the energy in the electron-positron plasma is then 
$\epsilon E_{\rm flow} \sim 8 \times 10^{50}$ ergs. The relatively wide 
variation in fluence and luminosity of GRBs observed so far should mainly 
be due to a distribution of the opening angle $\theta$ appearing in 
Eq.~(\ref{Eiso2}). 

It has long been recognized that GRB durations are distributed bimodally, 
with short bursts clustered around $\sim 0.2$ s and long events clustered 
around $\sim 20$ s (Mazets et al. 1981; Mao, Narayan \& Piran 1994). 
Currently, afterglows have 
been observed only from long GRBs, so that the distances and the nature of 
short GRBs are completely uncertain. It is very interesting that our model
also provides a natural explanation for the existence of these short 
bursts, since the progenitors here are millisecond magnetars. The advantage 
of millisecond magnetars to explain the bimodal duration distribution of 
GRBs has been discussed by Usov (1992) and Yi \& Blackman (1998). The
key point is that there exists a critical rotating period ($P_{\rm cr}$) 
for pulsars. The critical period $P_{\rm cr}$ depends on neutron star 
mass and is $\sim 0.5$ --- 1.6 ms (Friedman 1983; Usov 1992). If a pulsar
rotates with a period smaller than $P_{\rm cr}$, instability arises inside 
the compact star so that gravitational radiation plays the major in 
braking the fast rotator. In this case, the spin-down timescale becomes
(Usov 1992)
\begin{equation}
\tau_{\rm GW} \approx 0.12 \left(\frac{\varepsilon}{0.1}
\right)^{-2} \cdot \left(\frac{P}{0.5 \, {\rm ms}}\right)^4 \;\; {\rm s}, 
\;\;\;\;\;\; {\rm with} \;\; P < P_{\rm cr},
\label{tau8}
\end{equation}
where $\varepsilon$ is the equatorial ellipticity of the neutron star and 
is typically a few times 0.1 .   
Abundant high energy particles emitted during this quick deceleration phase 
can generate the observed short GRBs (Usov 1992; Yi \& Blackman 1998). 
A reasonable inference of this model is that short GRBs might also be 
highly collimated. The testing of such collimation should be an interesting 
goal in future observations of short GRBs. Furthermore, the observed number
of short GRBs relative to that of long GRBs might give us some hints on the 
distribution of the initial periods of magnetars at birth. 

\section{Conclusion and Discussion}

The connection between GRBs and neutron star kicks is a natural deduction 
from momentum conservation (Dar \& Plaga 1999; Dar 1999). 
Here we suggest that the neutron star in this mechanism should be 
a high speed one, with velocity larger than $\sim 1000$ km/s. We 
have shown that the model can naturally satisfy many of the observational 
constraints on the central engine of GRBs. For example, it well explains 
the energetics, the collimation, the event rate, the ultra-relativistic 
motion, the light curve variability in $\gamma$-rays, the bimodal 
distribution of durations, the narrowly clustered intrinsic energy, and 
the association of GRBs with supernovae and star forming regions. 
We also discuss the implications of this model on the neutron star kick 
mechanism, and suggest that the high kick speed is most likely acquired 
due to the electromagnetic rocket effect of a millisecond 
magnetar with an off-centered magnetic dipole. 

In all our discussion in the previous sections, we have assummed that 
a single recoiling outflow is responsible for the kick of the pulsar. 
However, Dar et al. (Dar \& Plaga 1999; Dar 1999) 
have pointed out that in realistic case two antiparallel jets might
be ejected by the neutron star. Then it is the momentum imbalance in 
these two jets that is responsible for the large kick velocity. In this 
case, the energy in our Eq.~(\ref{Eflow1}) is only a lower limit of the 
dominant jet. An interesting consequence of this picture is that in some
cases it might be the weaker jet, not the dominant one, that is pointing
toward us. Since the energy is much less now, it is very likely that we would
observe a failed gamma-ray burst (FGRB), i.e., a relativistic outflow with
the Lorentz factor $1 \ll \gamma \ll 100$ --- 1000 (Huang, Dai \& Lu 2002). 
Huang, Dai \& Lu (2002) have suggested that such FGRBs might give birth to
the so called X-ray flashes, a kind of GRB-like X-ray transients that were 
identified very recently (Strohmayer et al. 1998; Frontera et al. 2000; 
Kippen et al. 2001; Barraud et al. 2003). Totani (2003) further pointed out
clearly that FGRBs might usually be associated with supernovae. 

\acknowledgments

We thank the anonymous referee for valuable suggestions.  
YFH thanks A. De R\'ujula, A. Dar, and S. Dado for helpful communication.
This research was supported by the Special Funds for Major State
Basic Research Projects, the National Natural Science Foundation
of China, the Foundation for the Author of National Excellent
Doctoral Dissertation of P. R. China (Project No: 200125), the
National 973 Project (NKBRSF G19990754), and a RGC grant of
Hong Kong SAR.




\end{document}